%% file: paper_master.tex
\documentclass[useAMS,usenatbib,a4paper]{mn2e}

\usepackage[left=2cm,top=2cm,right=2cm,bottom=2cm]{geometry}
\usepackage{amssymb,epsfig,color}
\usepackage{amsmath}
\usepackage{graphicx}
\usepackage[font=footnotesize]{subfig}
\usepackage{xspace}
\usepackage{hyperref}
\usepackage{mn2e-bst}

\input{commands.tex}
\newcommand{\cc}[1]{\multicolumn{1}{|c|}{#1}}

\title[Polarized light curve of GRB131030A]{Early-time polarized optical light curve of GRB\,131030A}
\author[O.\,G.~King et al.]{O.\,G.~King$^{1}$\thanks{E-mail:ogk@astro.caltech.edu}, 
D.~Blinov$^{2,7}$,
D.~Giannios$^{9}$,
I.~Papadakis$^{2,6}$,
E.~Angelakis$^{4}$,
\newauthor
M.~Balokovi\'{c}$^{1}$, 
L.~Fuhrmann$^{4}$, 
T.~Hovatta$^{1,8}$, 
P.~Khodade$^{3}$, 
S.~Kiehlmann$^{4}$,
\newauthor
N.~Kylafis$^{2,6}$,
A.~Kus$^{5}$, 
I.~Myserlis$^{4}$,
D.~Modi$^{3}$,
G.~Panopoulou$^{2}$, 
\newauthor
I.~Papamastorakis$^{2,6}$, 
V.~Pavlidou$^{6,2}$,
B.~Pazderska$^{5}$,
E.~Pazderski$^{5}$, 
\newauthor
T.\,J.~Pearson$^{1}$, 
C.~Rajarshi$^{3}$,
A.\,N.~Ramaprakash$^{3}$,
A.\,C.\,S.~Readhead$^{1}$, 
\newauthor
P.~Reig$^{6,2}$,
K.~Tassis$^{2,6}$,
J.\,A.~Zensus$^{4}$
\\
$^{1}$Cahill Center for Astronomy and Astrophysics, California Institute of Technology, 1200 E
California Blvd, MC 249-17, \\Pasadena CA, 91125, USA\\
$^{2}$Department of Physics and Institute of Theoretical \& Computational Physics, University of
Crete, PO Box 2208, \\GR-710 03, Heraklion, Crete, Greece\\
$^{3}$Inter-University Centre for Astronomy and Astrophysics, Post Bag
4, Ganeshkhind, Pune - 411 007, India\\
$^{4}$Max-Planck-Institut f\"{u}r Radioastronomie, Auf dem H\"{u}gel
69, 53121 Bonn, Germany\\
$^{5}$Toru\'{n} Centre for Astronomy, Nicolaus Copernicus University, Faculty of
Physics, Astronomy and Informatics, \\Grudziadzka 5, 87-100 Toru\'{n}, Poland \\
$^{6}$Foundation for Research and Technology - Hellas, IESL, Voutes, 7110 Heraklion, Greece \\
$^{7}$Astronomical Institute, St. Petersburg State University,Universitetsky pr. 28,
Petrodvoretz, 198504 St. Petersburg, Russia \\
$^{8}$Aalto University Mets\"ahovi Radio Observatory, Mets\"ahovintie
114, 02540 Kylm\"al\"a, Finland \\
$^{9}$Department of Physics and Astronomy, Purdue University, 525 Northwestern Avenue, West Lafayette, IN 47907, USA
}

\begin{document}

\date{Accepted XXX. Received YYY; in original form ZZZ}

\pagerange{\pageref{firstpage}--\pageref{lastpage}} \pubyear{2014}

\maketitle

\label{firstpage}

\begin{abstract}
We report the polarized optical light curve of a gamma-ray burst afterglow obtained using the RoboPol instrument. Observations began 655~seconds after the initial burst of gamma-rays from GRB\,131030A, and continued uninterrupted for 2~hours. The afterglow displayed a low, constant fractional linear polarization of $p = (2.1 \pm 1.6)\,\%$ throughout, which is similar to the interstellar polarization measured on nearby stars. The optical brightness decay is consistent with a forward-shock propagating in a medium of constant density, and the low polarization fraction indicates a disordered magnetic field in the shock front. This supports the idea that the magnetic field is amplified by plasma instabilities on the shock front. These plasma instabilities produce strong magnetic fields with random directions on scales much smaller than the total observable region of the shock, and the resulting randomly-oriented polarization vectors sum to produce a low net polarization over the total observable region of the shock.
\end{abstract}


\section{Introduction} \label{sec:introduction}

Gamma-ray burst (GRB) afterglows are usually attributed to the synchrotron emission from a shock or jet propagating through the circumburst medium. The observed emission is thought to be the combination of the forward shock and a reverse shock that propagates backward into the flow \citep{1999PhR...314..575P,2003ApJ...595..950Z}, with the reverse shock dominating at early times. The light from the reverse shock might be highly linearly polarized if ordered magnetic fields thread the ejecta \citep{2003ApJ...594L..83G,2003MNRAS.346..540L,2004A&A...422..121L}, while the polarization of the forward shock depends of the circumburst magnetic field \citep{2012ApJ...752L...6U}.



The early-time polarized optical GRB afterglow emission has been measured five times. \citet{2007Sci...315.1822M} measured a $2\sigma$ upper limit of 8\,\% on the linear polarization 203\,s after the GRB event for GRB\,060418. They interpreted this relatively-low polarization level as ruling out the presence of a large-scale ordered magnetic field. 
The next measurement of the early-time afterglow polarization was made by \citet{2009Natur.462..767S} of GRB\,090102 160.8\,s after the GRB. They, by contrast, measured a level of $(10 \pm 1)\,\%$, which they interpreted as coming from the reverse shock. GRB\,110205A was measured by \citet{2011ApJ...743..154C} to have a $3\sigma$ upper limit of 16\,\% 243\,s after the BAT trigger time. A later measurement 56\,min after the trigger time found a polarization level of $3.6^{+2.6}_{-3.6}\,$\% ($2\sigma$ confidence levels). They excluded the zero-polarization hypothesis at a 92\,\% confidence level, supporting a reverse plus forward-shock scenario. \citet{2012ApJ...752L...6U} measured the optical polarization afterglow of GRB\,091208B from 149 to 706\,s after the burst trigger and found a linear polarization level of $(10.4\pm2.5)\,\%$. At the time of the measurement the optical light curve exhibited a power-law decay (index of $-0.75 \pm 0.02$), which they interpreted as the signature of the forward shock synchrotron emission.

Most recently, \citet{2013Natur.504..119M} obtained multiple measurements of the early-time optical polarization light curve of GRB\,120308A, making this the first measurement of the temporal evolution of the early-time polarized optical afterglow emission. They began observing the GRB afterglow 240\,s after the GRB trigger and monitored it for $\sim 10$\,min, during which time the fractional polarization dropped from $28^{+4}_{-4}$\,\% to $16^{+5}_{-4}$\,\%.

\section{RoboPol observations of GRB\,131030A} \label{sec:robopol_observations}

The RoboPol project operates a four-channel imaging polarimeter on the 1.3\,m telescope at the Skinakas Observatory in Crete, Greece\footnote{\url{http://skinakas.physics.uoc.gr/}}. The RoboPol instrument measures the Stokes parameters $I$, $q=Q/I$, and $u=U/I$, simultaneously in a single exposure. It is used to monitor the optical linear polarization of blazars \citep{robopol_survey}, and observations are performed by an automated control system \citep{robopol_pipeline} that is capable of responding to target-of-opportunity (TOO) events such as GRBs.

At 20:56:18 UT on 2013 October 30 the Swift Burst Alert Telescope (BAT, \citealt{2004SPIE.5165..175B}) triggered and located GRB\,131030A. The afterglow was located at 23$^{\rm h}$00'16.13'', $-$05$^{\circ}$22'05.1'' (J2000) by the Swift Ultraviolet/Optical Telescope (UVOT, GCN\#15402\footnote{\url{http://gcn.gsfc.nasa.gov/gcn3/15402.gcn3}}). The duration over which 90\,\% of the 15 -- 350\,keV GRB photons were collected, $T_{90}$, was $41.1 \pm 4.0\,$s and it had a fluence in the 15 -- 150\,keV band of $2.93 \pm 0.04 \times 10^{-5}$\,erg\,cm$^{-2}$ (GCN\#15456\footnote{\url{http://gcn.gsfc.nasa.gov/gcn3/15456.gcn3}}). The GRB occurred at a redshift of 1.293 -- 1.295 (GCN\#15407\footnote{\url{http://gcn.gsfc.nasa.gov/gcn3/15407.gcn3}} and GCN\#15408\footnote{\url{http://gcn.gsfc.nasa.gov/gcn3/15408.gcn3}}) and had an isotropic energy release of $E_{\rm iso} = (3.0 \pm 0.2) \times 10^{53}$\,erg (GCN\#15413\footnote{\url{http://gcn.gsfc.nasa.gov/gcn3/15413.gcn3}}).

\begin{figure}
 \centering
 \includegraphics[width=0.5\textwidth]{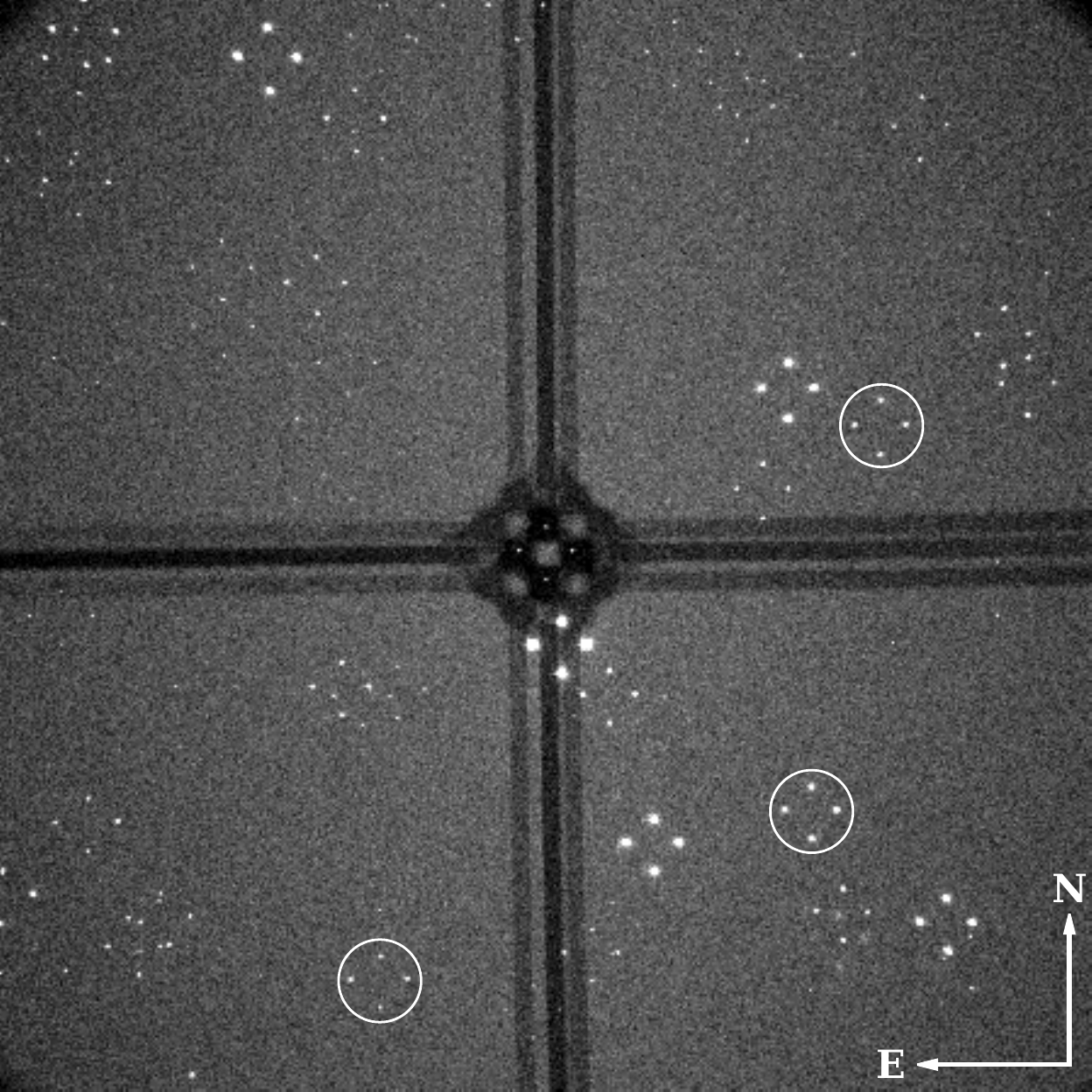}
\caption{A raw frame showing the characteristic four-spot pattern from the RoboPol instrument. The GRB is located in the low-background central area (four spots against a dark background arranged in a cross). The three reference stars used to provide relative photometry are circled.
The relative Stokes parameters $q$ and $u$ are obtained by taking ratios of the flux in pairs of spots.
}
 \label{fig:example_image}
\end{figure}

The RoboPol control system automatically responded to the GRB notification by interrupting the regular observing schedule and slewing to the location of the GRB afterglow. 
The telescope operator identified the afterglow and began taking exposures in the Johnson-Cousins $R$-band at 2013 October 10 21:07:13 UT, 655\,s after the GRB trigger. We continued monitoring the GRB afterglow in a series of exposures until it set below our observing horizon about 2\,h after the GRB, adjusting the exposure time as the afterglow faded. A typical image from the series of RoboPol exposures is shown in Figure~\ref{fig:example_image}.

The data were reduced using both the Aperture Photometry Tool \citep{APT} and the RoboPol pipeline \citep{robopol_pipeline}, and were calibrated using the RoboPol instrument model. Relative photometry was performed using three field sources (circled in Figure~\ref{fig:example_image}), with $R$-band magnitudes taken from the USNO-B1.0 photometric catalog \citep{USNOB}. The measurements from each exposure are given in Table~\ref{tab:robopol_data}.

\begin{table}
\centering
\caption{RoboPol data for GRB131030A. $t_m$ is the middle of the exposure time in the observer frame, in minutes since 2013 October 30 20:56:18 UT. $t_e$ is the exposure duration. The uncertainties in the $R$-band magnitude, $\sigma_R$, are dominated by a systematic uncertainty from the relative photometry fit.}
\label{tab:robopol_data}
\begin{tabular}{|r|r|r|r|r|r|r|r|} \hline
\cc{$t_m$} & \cc{$t_e$} & \cc{$p$} & \cc{$\sigma_p$} & \cc{$\chi$} & \cc{$\sigma_{\chi}$} & \cc{$R$} & \cc{$\sigma_R$} \\ 
\cc{min}      & \cc{s}  	& \cc{\%}  &  		& \cc{deg} & 			    & \cc{mag} &  \\ \hline \hline
\input{figures/table.tex} \hline
\end{tabular}
\end{table}
 


\begin{figure}
 \centering
 \includegraphics[width=0.5\textwidth]{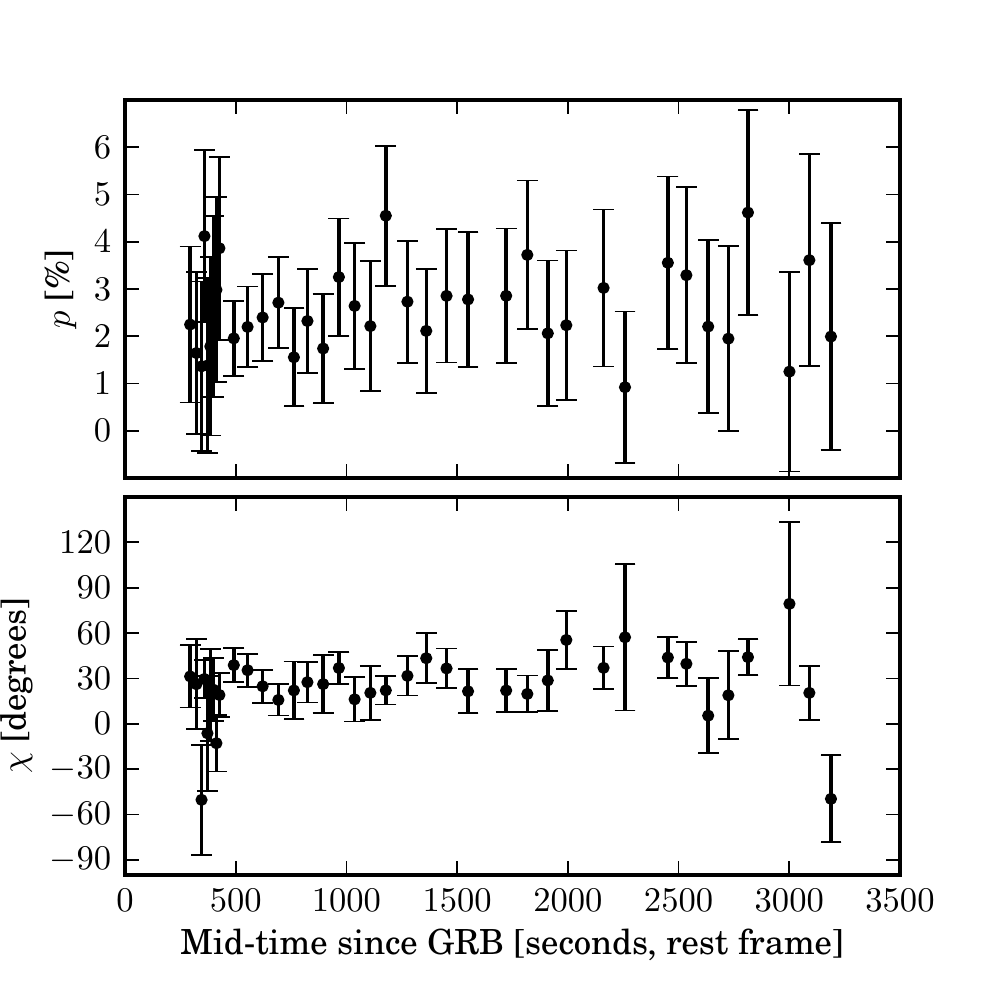}
 \caption{The $R$-band linearly polarized light curve of GRB\,131030A as measured by the RoboPol instrument. The fractional linear polarization $p$ is listed as a percentage of the total light, and the EVPA $\chi$ is given in degrees.}
 \label{fig:robopol_p_lightcurve}
\end{figure}

The linearly polarized light curve for the GRB afterglow is shown in Figure~\ref{fig:robopol_p_lightcurve}. The polarization measurements have not been debiased, as most data points have $p/\sigma_p > \sqrt{2}$, the threshold at which debiasing is usually applied \citep{robopol_survey}. The linear polarization behavior of the afterglow appears to remain constant throughout the 2~hour observing period. The mean polarization percentage from our data is $p = (2.1 \pm 1.6)\,\%$, and the mean polarization angle is $\chi = 27^{\circ} \pm 22$. 

\begin{figure}
 \centering
 \includegraphics[width=0.45\textwidth]{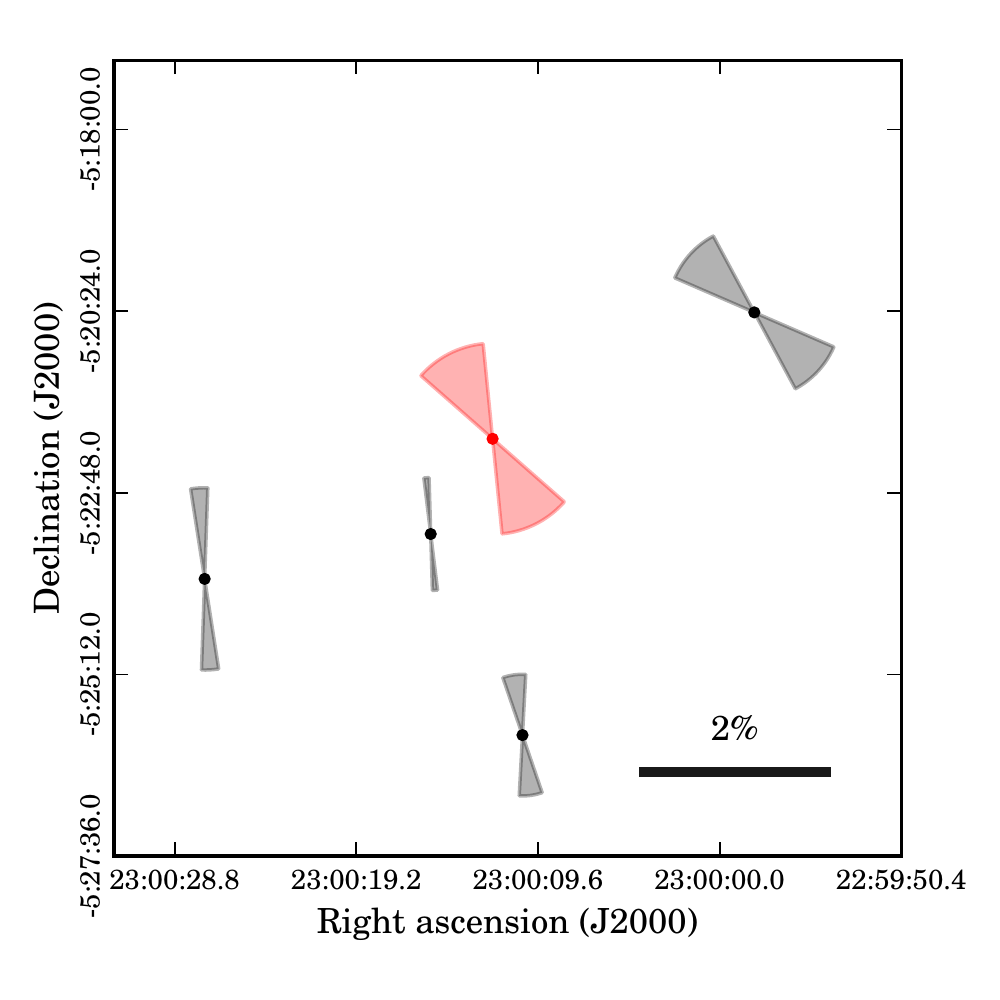}
\caption{
A polarization vector map of the field around GRB\,131030A. Seperate observations were made of four field stars around GRB\,131030A, indicated by the grey wedges. GRB\,131030A is indicated by the light-red wedges. The diameter of the wedge is proportional to the polarization percentage, while the angle subtended indicates the $1\sigma$ polarization angle.
}
 \label{fig:pol_map}
\end{figure}

\section{Is the polarization intrinsic?}

The measured polarization of the GRB might be due to interstellar extinction in our Galaxy. We can estimate the expected level of induced polarization in the direction of the GRB from the level of Galactic extinction using the standard empirical relation from \citet{1975ApJ...196..261S}. According to the NASA Extragalactic Database the extinction in the direction of the GRB is $A_B = 0.208$ and $A_V = 0.157$, which gives $E(B-V) = 0.051$\,mag \citep{2011ApJ...737..103S}. The resulting level of stellar polarization is $P_{\rm max} \leq 9.0 E(B-V)$, i.e., $\sim 0.5$\,\%, though this method is approximate.

To obtain a more accurate estimate of the scale of the interstellar scattering effect we measured the linear polarization of four field stars around GRB\,131030A in a separate series of exposures. 
We show in Figure~\ref{fig:pol_map} a polarization vector map of the GRB and the field sources.
The mean polarization fraction for the field sources is $(1.66 \pm 0.43)\,\%$, and the polarization vectors are well-aligned, indicating an ordered magnetic field in the absorbing interstellar medium (ISM). 

The high level of polarization of the field sources around GRB\,131030A implies that the measured polarization is dominated by interstellar extinction rather than the intrinsic polarization of the GRB afterglow.

\section{Interpretation} \label{sec:interpretation}

The GRB occurred at a redshift of 1.294, so in the rest-frame we started observing $655/(1+z) = 285$\,s after the GRB event, which corresponds to about $16 \times T_{90}$.
This is about 3--5 times longer than the time when the 5 early-time optical polarimetric observations we mention in the Introduction started; Table~\ref{tab:all_polarization_measurements} summarizes these data.

\begin{table*}
\centering
\caption{Measurements of the optical polarization of the early GRB afterglow emission. The times are in the rest-frame, i.e., have been corrected for redshift. The interpretation column indicates whether the authors interpreted the optical emission as being dominated by either the forward or reverse shock, or whether they contribute approximately equally.}
\label{tab:all_polarization_measurements}
\begin{tabular}{|l|l|l|l|l|} \hline
Name 			& $t_{\rm start}$ (s) 	& $t_{\rm exp}$ (s) & Polarization & Interpretation \\ \hline\hline
GRB\,120308A \citep{2013Natur.504..119M}	& 90 	& 135 	& $28-15\,\%$ 		& Reverse shock \\
GRB\,090102 \citep{2009Natur.462..767S}		& 63 	& 24 	& $(10 \pm 1)\,\%$ 	& Reverse shock \\
GRB\,110205A \citep{2011ApJ...743..154C}	& 76 	& ? 	& $<16\,\%$ 		& Reverse shock \\
GRB\,060418 \citep{2007Sci...315.1822M}		& 82 	& 12 	& $<8\,\%$ ($2\sigma$) 	& Both \\
GRB\,091208B \citep{2012ApJ...752L...6U}	& 72 	& 551 	& $(10.4 \pm 2.5)\,\%$ 	& Forward shock \\
GRB\,131030A (this work)			& 285 	& 2894 	& $< 2\,\%$ 		& Forward shock \\ \hline
\end{tabular}
\end{table*}

\begin{figure}
\centering
\includegraphics[width=0.5\textwidth]{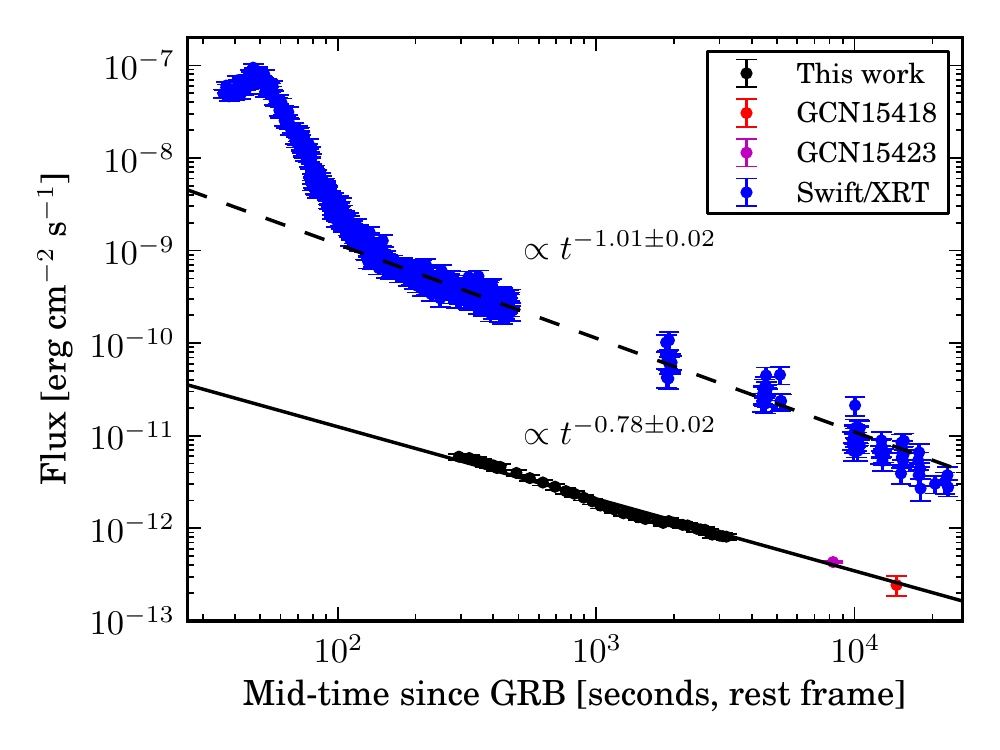}
\caption{The optical and X-ray light curves GRB\,131030A, including measurements published in GCNs, with the best-fit power-law curves.}
\label{fig:light_curves_with_fits}
\end{figure}

The X-ray Telescope (XRT; \citealt{2005SSRv..120..165B}) started observing the GRB field 78.4\,s\footnote{\url{http://gcn.gsfc.nasa.gov/gcn3/15402.gcn3}} after the trigger. The XRT light curve\footnote{\url{http://www.swift.ac.uk/xrt_curves/00576238/flux.qdp}} is shown in Figure~\ref{fig:light_curves_with_fits}. At early times, the X-ray light curve brightens until $\sim 50\,$s (rest frame) after the burst, and then the X-ray afterglow decays steeply until $\sim 150\,$s. At later times, coincident with the RoboPol observations, the X-ray light curve declines as a single power law $\propto t^{-1.01 \pm 0.02}$.
The RoboPol optical light curve is also plotted in Figure~\ref{fig:light_curves_with_fits}. The optical flux declines also as a single power law $\propto t^{-0.78 \pm 0.02}$ (we have included in the optical band light curve later $R$-band photometry from GCN circulars \#15418\footnote{\url{http://gcn.gsfc.nasa.gov/gcn3/15418.gcn3}} and \#15423\footnote{\url{http://gcn.gsfc.nasa.gov/gcn3/15423.gcn3}}). 

Since we observe a single, power-law  decline from $\sim 5$ up to $\sim 55\,$min (rest-frame) after the burst, both in the optical and X-rays, the simplest explanation is that a single emitting component is responsible for the observed emission in both bands. This is most probably the forward shock propagating in the external medium, and we observe the synchrotron emission from this shock. 
The X-ray decline is consistent with the fast-cooling afterglow from a shock that has a power-law distribution of electron energies with a spectral index of $p_E = 2.01 \pm 0.03$ ($F_X \propto t^{(2-3p_E)/4}$, \citealt{2002ApJ...568..820G}). 
If the GRB ambient density profile is similar to the ISM we would expect an optical light curve that evolves as $\propto t^{-3/4}$ while in an environment with a stellar-wind density profile the light curve would evolve as $\propto t^{-5/4}$. The measured optical power-law index of $0.78 \pm 0.02$ implies that the medium surrounding the GRB has a constant density with a profile similar to the ISM. 
The blast decelerates in a constant density medium and the cooling synchrotron break is between the optical and the X-ray bands $\sim 5-55\,$min after the burst. The observed flux in the X-ray and optical bands is consistent with this model if a fraction $\epsilon_e\sim 0.1 $ of the dissipated energy goes into non-thermal electrons, a fraction $\epsilon_B\sim 3\times 10^{-4}$ goes into amplifying the magnetic field while the ambient density of the circumburst material is $n\sim 1$\,cm$^{-3}$. These values are similar to those inferred in other bursts (e.g., \citealt{2014ApJ...785...29S}).    

In general, the GRB afterglow emission is believed to consist mainly of the reverse and forward shock emission and other possible components (such as radiation related to jet reactivation). The reverse shock emission may dominate at early times, i.e., comparable, or a few times longer that the duration $T_{90}$ of the burst \citep{1999ApJ...513..669K,2009A&A...494..879M,2010MNRAS.407.2501M} and may be strongly polarized, as in  \citet{2013Natur.504..119M,2009Natur.462..767S}. At later times, the forward shock is the primary candidate for emission, and its polarization may be much weaker as observed by \citet{2007Sci...315.1822M}. Our results support this view, indicating a disordered magnetic field in the shock front as it propagates through the ambient medium around GRB\,131030A. 

This result supports suggestions that the magnetic field is amplified by plasma instabilities on the shock front, which would produce strong magnetic fields with random directions, on scales much smaller than the total observable region of the shock \citep{1999ApJ...526..697M}. On the other hand, \citet{2012ApJ...752L...6U} observed a strong polarization signal of $\sim 10\,\%$ from the early afterglow of GRB\,091208B, when the observed emission was also dominated by the forward shock emission. Their observation started $\sim 72\,$s after the burst, and lasted for $\sim 550\,$s (in the source rest frame). They measured an $R$-band flux decaying as $t^{-0.75}$, with the X-ray flux initially decaying as $t^{-0.18}$ and later steepening to $t^{-1}$.
The first $\sim 200\,$s of our observations overlap with the end of their observations in rest-frame time, during which time both the optical and X-ray light curve decay rates are very similar.  Therefore, if the same mechanism operates in all GRBs, then a very fast decline in optical polarization must take place, indicating a fast change in the mechanism that amplifies the strong magnetic fields in the jet of these sources. On the other hand, these mechanisms may not be the same in all GRBs. More optical polarization data from different GRBs, and on long time scales, are needed in order to understand better the magnetic field structure in GRBs.

\section*{Acknowledgments}

The RoboPol project is a collaboration between Caltech in the USA,
MPIfR in Germany, Toru\'{n} Centre for Astronomy in Poland, the University of
Crete/FORTH in Greece, and IUCAA in India.
The U. of Crete group acknowledges support by the ``RoboPol'' project, which is implemented under
the ``Aristeia'' Action of the  ``Operational Programme Education and Lifelong Learning'' and is
co-funded by the European Social Fund (ESF) and Greek National Resources, and by the European
Comission Seventh Framework Programme (FP7) through grants PCIG10-GA-2011-304001 ``JetPop'' and
PIRSES-GA-2012-31578 ``EuroCal''.
This research was supported in part by NASA grant NNX11A043G and NSF grant AST-1109911, and by the
Polish National Science Centre, grant number 2011/01/B/ST9/04618.
K.\,T. acknowledges support by the European Commission Seventh Framework Programme (FP7) through
the Marie Curie Career Integration Grant PCIG-GA-2011-293531 ``SFOnset''.
M.\,B. acknowledges support from the International Fulbright Science and Technology Award.
I.\,M. and S.\,K. are supported for this research through a stipend from the International Max Planck Research School (IMPRS) for Astronomy and
Astrophysics at the Universities of Bonn and Cologne.
T.\,H. was supported by the Academy of Finland project number 267324.

This research made use of Astropy, \url{http://www.astropy.org}, a community-developed core Python
package for Astronomy \citep{astropy}.

This research has made use of the NASA/IPAC Extragalactic Database (NED) which is operated by the Jet Propulsion Laboratory, California Institute of Technology, under contract with the National Aeronautics and Space Administration.


\bibliographystyle{mn2e}
\bibliography{bibliography_manual,bibliography_export}

\label{lastpage}

\end{document}

%% file: figures/table.tex
11.25 & 20 & 2.25 & 1.65 & 31.3 & 20.6 & 15.94 & 0.08 \\
12.33 & 20 & 1.65 & 1.72 & 26.2 & 29.8 & 15.97 & 0.08 \\
13.22 & 20 & 1.36 & 1.80 & 129.7 & 36.4 & 16.05 & 0.08 \\
13.72 & 20 & 4.12 & 1.82 & 29.6 & 12.4 & 16.09 & 0.08 \\
14.23 & 20 & 1.38 & 1.85 & 173.7 & 38.0 & 16.12 & 0.07 \\
14.73 & 20 & 1.79 & 1.89 & 19.1 & 30.3 & 16.15 & 0.08 \\
15.28 & 20 & 2.63 & 1.91 & 22.5 & 20.8 & 16.16 & 0.07 \\
15.78 & 20 & 2.98 & 1.96 & 167.1 & 18.7 & 16.22 & 0.08 \\
16.30 & 20 & 3.86 & 1.94 & 19.0 & 14.4 & 16.20 & 0.07 \\
18.78 & 120 & 1.96 & 0.80 & 38.8 & 11.2 & 16.38 & 0.07 \\
21.17 & 120 & 2.20 & 0.85 & 35.4 & 10.9 & 16.51 & 0.07 \\
23.77 & 120 & 2.40 & 0.92 & 24.8 & 10.8 & 16.62 & 0.07 \\
26.48 & 120 & 2.71 & 0.97 & 15.8 & 10.4 & 16.75 & 0.07 \\
29.17 & 120 & 1.56 & 1.04 & 22.0 & 19.1 & 16.87 & 0.08 \\
31.50 & 120 & 2.32 & 1.10 & 27.5 & 13.4 & 16.94 & 0.07 \\
34.22 & 120 & 1.74 & 1.16 & 26.2 & 19.1 & 17.06 & 0.07 \\
36.93 & 120 & 3.25 & 1.24 & 36.9 & 10.5 & 17.16 & 0.07 \\
39.65 & 120 & 2.64 & 1.33 & 16.2 & 14.7 & 17.32 & 0.07 \\
42.37 & 120 & 2.21 & 1.38 & 20.5 & 17.8 & 17.39 & 0.07 \\
45.03 & 120 & 4.55 & 1.48 & 22.1 & 9.3 & 17.45 & 0.08 \\
48.77 & 180 & 2.73 & 1.29 & 31.7 & 13.0 & 17.55 & 0.07 \\
52.02 & 180 & 2.11 & 1.32 & 43.4 & 16.5 & 17.53 & 0.07 \\
55.52 & 180 & 2.85 & 1.41 & 36.6 & 13.2 & 17.65 & 0.07 \\
59.23 & 180 & 2.78 & 1.42 & 21.5 & 14.5 & 17.69 & 0.07 \\
65.82 & 180 & 2.86 & 1.42 & 22.0 & 14.2 & 17.72 & 0.08 \\
69.47 & 180 & 3.72 & 1.57 & 19.7 & 12.2 & 17.84 & 0.08 \\
73.00 & 180 & 2.06 & 1.54 & 28.7 & 20.3 & 17.68 & 0.08 \\
76.20 & 180 & 2.23 & 1.58 & 55.4 & 19.1 & 17.71 & 0.08 \\
82.63 & 180 & 3.02 & 1.66 & 37.0 & 14.1 & 17.79 & 0.07 \\
86.33 & 180 & 0.92 & 1.60 & 57.2 & 48.4 & 17.79 & 0.08 \\
93.73 & 180 & 3.56 & 1.82 & 43.8 & 13.7 & 17.90 & 0.08 \\
96.93 & 180 & 3.29 & 1.86 & 39.7 & 14.6 & 17.92 & 0.08 \\
100.70 & 180 & 2.21 & 1.83 & 5.4 & 24.9 & 17.92 & 0.08 \\
104.18 & 180 & 1.95 & 1.96 & 18.9 & 29.1 & 18.09 & 0.08 \\
107.57 & 180 & 4.62 & 2.17 & 44.1 & 11.9 & 18.10 & 0.08 \\
114.73 & 180 & 1.25 & 2.11 & 79.3 & 53.9 & 18.12 & 0.08 \\
118.17 & 180 & 3.61 & 2.24 & 20.4 & 17.7 & 18.27 & 0.08 \\
121.90 & 180 & 1.99 & 2.40 & 130.4 & 28.8 & 18.12 & 0.08 \\